\documentclass[12pt,a4paper]{article}
\usepackage{natbib}
\usepackage{times} 
\usepackage{indentfirst}
\usepackage{xurl}
\usepackage{hyperref}
\usepackage{longtable}
\usepackage{multirow}  
 
\usepackage{makecell}
\begin{document}
\begin{center}

{\large{\bf Integrating GenAI in Filmmaking:\\
\vspace{0.3cm}
From Co-Creativity to Distributed Creativity} }

\vskip0.5\baselineskip{\bf  {Pierluigi Masai}$^{1}$,  Lorenzo Carta$^{2}$, and Mateusz Miroslaw Lis${^3}$ }

\vskip0.5\baselineskip{\em$^{1}$University of Trieste, Department of Mathematics, Informatics and Geosciences, Trieste, Italy, pierluigi.masai@phd.units.it 
\\$^{2}$University of Trieste, Department of Political and Social Sciences, Trieste, Italy, lorenzo.carta2@phd.units.it
\\$^{3}$SSML Ciels, Padova, Italy, mateuszmiroslaw.lis@ciels.it
}\\

\end{center}

\noindent
\section*{Abstract}

The integration of Generative AI (GenAI) into audio-visual production is often presented as a radical break from past traditions. However, through a sociomaterial and historical lens, this paper argues that GenAI represents a new development in the long-standing negotiation between creative labor and technological possibilities. Moving beyond the limiting framework of human-machine \textit{co-creativity}, we adopt an STS-based approach to investigate \textit{creativity in the making} within the Filmmaking industry. We analyze Filmmaking as a distributed process where agency is shared across diverse human experts and non-human actors, showing how technological innovations have historically reconfigured Filmmaking practices long before the advent of AI. The article introduces an analytical taxonomy of GenAI techniques to illustrate how these technologies do not merely “assist” but can actively reconfigure professional roles, production temporalities, and film aesthetics. By linking sociomaterial configurations to aesthetic outcomes, this reframing suggests that AI technologies in Filmmaking should be seen as mediators that could enable new aesthetic possibilities by blurring the boundaries of traditional filmmaking workflows.

\section*{Introduction\footnote{All authors developed the concept for this work, conducted the literature review, and provided editorial revisions. LC drafted sections 1 and 2; PM drafted section 3; MML drafted sections 4.}}

Recent developments in Generative AI (GenAI) systems have raised both preoccupation and curiosity regarding their actual capacities and the long-term consequences on creative industries and practices. The current debate is heavily characterized by a tension between techno-optimist promises of “democratized” creativity and fears concerning labor displacement and the erosion of authorship. Dominant media narratives often frame GenAI as a tool that “liberates” creativity from the friction of material realization, decoupling the “idea” from the human labor required to execute it (\citealp{caramiaux_generative_2025}; \citealp{celis_bueno_enduring_2025}).

However, these expectations, whether utopian or dystopian, frequently lack grounding in empirical observation and fail to consider the actual, situated practices of creative professionals. Furthermore, the perceived novelty of these changes is often overstated. The current discourse reactivates a historical tension between cultural production and technology that has emerged with every major technological shift, from the daguerreotype to digital technologies. By positioning GenAI as an unprecedented force, it is often overlooked how creative labor has always been a negotiation with technological constraints (\citealp{celis_bueno_enduring_2025}). On the other end, a common rhetoric frames AI as a servile “co-creative assistant” while obscuring the profound reconfigurations of labor and the material reality of the production (\citealp{chow_cloak_2025}).
The shortcomings of these narratives are reflected also in the current debate on “computational co-creativity.” Recently, the discourse surrounding AI and creativity has shifted from debating the autonomous capabilities of machines toward an analysis of human-AI interactions. This transition toward computational co-creativity emphasizes the collaboration between human and artificial agents in constructing shared artifacts (\citealp{rafner_computational_2025}). We hold this approach to be limiting as it often overlooks the situated interactions among the diverse actors, both human and technological, involved in the creative process, relegating such elements to the background of the AI-human interaction.

To move beyond these narratives and approaches, the analysis must ground the discussion in appropriate theoretical reflections and empirical observation, focusing on how these technologies are actually integrated into specific creative practices. With this paper, we aim to reframe the issue of creativity and GenAI through the lens of Science and Technology Studies (STS), integrated with insights from media production studies and film studies. Adopting STS methodologies to study “creativity in the making,” we investigate creative production as a distributed and heterogeneous process, where agency emerges from the interaction between human and non-human actors. From this perspective, GenAI is not an exceptional “co-creator” but a new mediator entering the complexity of creative labour.

To make our case we focus our attention on audio-visual production, more specifically on Filmmaking,\footnote{We capitalize Filmmaking as a way of meaning “the filmmaking process” which amounts to all the different phases, techniques, equipment and people needed, from pre-production to post-production and distribution.} i.e., the process of creating a motion picture. The case of Filmmaking is convenient to the present discussion for at least two reasons. First, Filmmaking typically involves many people with very different and specific expertise, which makes the case for distribution within the creative process, a distribution that is not just among human actors but among tools and equipment as well. Second, GenAI systems have raised much preoccupation and attention specifically for their capacities in generating images and sounds, which suggest their use may significantly redefine the entire Filmmaking industry other than challenging its very basis. We claim that the most relevant use of AI systems in Filmmaking is to be understood through the lens of sociomateriality, focusing on how they may reconfigure current practices

This approach to Filmmaking allows us to bridge two divergent perspectives on GenAI and creative practices (\citealp{chow_generative_2026}): an industrial approach, centered on the material conditions and labor processes of workers, and an artistic approach, focused on the aesthetics and symbolic meanings of art practices. As we will demonstrate, the aesthetic qualities of a film—its style, form, and narrative texture—do not emerge from isolated authorial choices, but are the emergent effects of the distributed practices and material conditions of its production, which can also include GenAI systems.

In order to achieve these aims, §\ref{Sec_1} addresses the notion of creativity and co-creativity in relation with computational systems, briefly outlining the main approaches within the theoretical debate. In §\ref{Sec_2} we critique the most prominent frameworks in the AI co-creativity discourse and deconstruct four prevalent assumptions that limit current understanding of AI integration into creative practices. Building on this critique, we establish a new set of methodological orientations for studying creativity in the making, grounded in the notion of distributed creativity. In §\ref{Sec_3} the new conceptual lens is applied to Filmmaking to demonstrate how it is a distributed process where agency is shared across professionals, spaces, and temporalities. By integrating film and media studies literature, we show how past technological innovations, long before GenAI, have historically reconfigured production workflows and film aesthetics, establishing the contextual precedent for the current integration of GenAI. Finally in §\ref{Sec_4}, drawing from the theoretical shifts identified in the previous phases, we develop a tentative taxonomy of AI techniques for Filmmaking. This will be used as an analytical tool to show the sociomaterial reconfiguration of practices, professional roles, and film aesthetics. The taxonomy distinguishes three main families of AI techniques. For each family a table that sums up the results of the analysis is presented in the appendix.

\section{Computational Creativity and Co-Creativity Systems}\label{Sec_1}

In recent years, there has been a growing interest in the field of machine learning and creativity, particularly concerning generative technologies (\citealp{franceschelli_creativity_2024}). To analyze these developments, it is crucial to distinguish between \textit{generative models} and \textit{GenAI systems}, as the latter represent a significantly broader and more complex entity. According to the OECD (\citeyear{oecd_explanatory_2024}), an AI model is a “core component” of the larger system. It acts as the internal mathematical engine—a logical or physical representation of data—producing inferences from input to data. The concept of an AI system is more encompassing because it functions as a complete operational architecture and it can be built using more than one model. This broader definition implies that a system includes not just the mathematical core, but also the interfaces, data processing layers, and the specific objectives—explicit or implicit—that govern its interaction with the world.

The recent advancements based on generative deep learning technologies, often collectively referred to as GenAI systems, encompass significant applications in both artistic and scientific domains (\citealp{ismayilzada_creativity_2024}). For instance, GenAI systems have been employed to produce poetry, images, musical pieces, or high-quality videos, where the creative output emerges not from the model in isolation, but through the operational dynamics of the entire system. These advancements have reinvigorated a longstanding debate on the possibility of machine creativity. This debate is a central theme in the field of \textit{computational creativity} (CC), which is defined as “the philosophy, science and engineering of computational systems which, by taking on particular responsibilities, exhibit behaviors that unbiased observers would deem to be creative” (\citealp[p. 1]{lamb_evaluating_2018}). The central objective of CC is twofold: (1) to design systems capable of exhibiting creative behavior and (2) to establish robust evaluation frameworks to assess their creativity.

One of the most widely used frameworks for analyzing creativity is the 4P’s model (Person, Process, Product, and Press), which provides a taxonomy for studying different dimensions of creativity. Originally proposed by Mel Rhodes in 1961, this framework was designed to categorize key strands in creativity research. Over time, it has evolved into a model for studying creativity, widely adopted in creativity studies (\citealp{glaveanu_rewriting_2013}). While early research in CC focused primarily on the process and the product, the field later adopted and expanded the 4P’s framework to overcome these limitations and provide a more comprehensive evaluation of creative systems (\citealp{jordanous_four_2016}). The 4P framework has been applied to analyze both human and CC, and in our case GenAI systems.

The 4P’s framework, when examined from a CC perspective, identifies four key dimensions of creativity: Person (or Producer), Process, Product, and Press (or Environment) (\citealp{jordanous_four_2016}; \citealp{lamb_evaluating_2018}; \citealp{carnovalini_computational_2020}). First, the “Person (or Producer)” perspective focuses on the inherent attributes that creative individuals, groups, or machines possess, such as traits, attitudes, and skills. Second, the “Process” dimension examines the cognitive activities an individual undertakes to be creative, including inspiration, idea development, and refinement. Third, the “Product” dimension refers to an artifact that is seen as creative or that is produced as a result of the creative process. This perspective focuses on the outputs or artifacts generated by creative systems, such as poems, music, or paintings. Finally, “The Press (or Environment)” refers to the bidirectional influence between the environment and the creator. This includes the cultural context, societal norms, available education, and training. Importantly, it also involves how the audience receives and judges creative work.\footnote{Many evaluation methods and techniques related to the 4P's perspectives have been proposed for computational and more recently for GenAI systems (\citealp{franceschelli_creativity_2024}; \citealp{ismayilzada_creativity_2024}). However, our objective is not to review these methods but rather to investigate the underlying assumptions that influence how creativity is ascribed to these systems.}

The discourse around AI and creativity is transitioning from debating whether a machine has an autonomous capacity for creativity to examining how collaborative human-AI interactions transform creative processes (\citealp{rafner_creativity_2023}; \citealp{moruzzi_artificial_2025}). This shift refocuses discussion on \textit{computational co-creativity systems}, defined as “a system involving at least one human agent and one artificially intelligent agent, collaborating with each other to build shared creative artifacts” (\citealp[p. 821]{rafner_computational_2025}). This recent area of research combines computational creativity debates and human-computer interaction (HCI) (\citealp{davis_ai_2025}). Co-creativity can take many forms that depend both on the design choices (e.g., prompt-based interactions) and the perceived roles of the users. Furthermore, to define co-creativity is something difficult to do because the possible roles attributable to GenAI systems in the creative process are multiple and highly context-dependent (\citealp{lin_ontology_2023}; \citealp{shanahan_role_2023}). Despite current uncertainties, GenAI is viewed as more than a mere “Creativity Support Tool” like Photoshop or Premiere. While those traditional tools primarily enhance human creativity, GenAI is expected to transition to an active collaborator, participating directly in the creative process and contributing autonomously to the final product (\citealp{palani_evolving_2024}; \citealp{rafner_computational_2025}).

Evaluating co-creativity is even more complex than computational creativity. Several frameworks have been proposed that are built on the previous discussion on CC and human computer interactions. The previous 4P’s dimensions could be used and expanded to include co-creative systems. The main difference between the evaluation of “autonomous” creative systems, which do not interact with humans during the creative processes, and co-creative systems is the central role of the interactions between the human and the AI system (\citealp{rezwana_designing_2022}).

Bown and Brown (\citeyear{filimowicz_interaction_2018}; \citealp{bown_beyond_2021}) have identified three broad interaction paradigms in CC. These paradigms range from Operation-based interaction, which involves the user directly manipulating the system's parameters; to Request-based interaction, like the prompt-based exchanges seen with GenAI; and finally, to Ambient-based interaction, where the AI system operates proactively or contextually in the background without waiting for an explicit user request. Several frameworks have been proposed for designing and evaluating these interactive aspects of co-creative systems (\citealp{kantosalo_five_2020}; \citealp{rezwana_designing_2022}; \citealp{lin_beyond_2023}; \citealp{davis_co-creative_2025}). §\ref{Sec_2} will highlight the common assumptions underlying these frameworks, and we argue that these core premises constitute the primary limitations of the current debate.

\section{From Co-Creativity to Distributed Creativity}\label{Sec_2}

In this section, we examine the main assumptions that underpin current research, offering a critical perspective informed by Science and Technology Studies (STS). While not an exhaustive account, our analysis focuses on what we see as the most limiting assumptions: a) the persistent description of creativity in cognitive terms; b) the narrow analytical focus on the human-AI dyad; c) the exceptional agency attributed to AI; d) the ontological approach that attempts at defining “what is creativity.” After demonstrating the shortcomings of these approaches, we introduce an alternative based on the concept of distributed creativity. We argue that this perspective is better equipped to capture the complex dynamics of creative production. Rather than furnishing another framework, our goal is to offer a set of methodological assumptions and orientations for the empirical study of creativity in the making.

\subsection{Common assumptions on Co-Creativity and AI}\label{Sec_2.1}

\subsubsection{Creativity is primarily described as a cognitive process}\label{Sec_2.1.1}

The prevailing frameworks within computational creativity describe the co-creative process primarily in cognitive terms, drawing heavily from cognitive science and psychological theories. For example, Rezwana and Maher (\citeyear{rezwana_designing_2022}) build their COFI framework on research into human-to-human collaboration and the cognitive theory of sense-making, which they then apply to human-AI systems. Within their framework, the creative product is defined in cognitive terms as “the idea or concept that is being created” (\citealp[p. 11]{rezwana_designing_2022}). Similarly, the Co-Creative Design Framework (CCDF) “incorporates concepts from enactive and embodied cognition, which emphasize the situated, embodied, and emergent nature of meaning-making and creative collaboration” (\citealp[p. 564]{davis_co-creative_2025}). The focus remains on cognitive phenomena like “ideas” and “concepts” rather than the broader materiality of creativity.

This cognitive framing tends to neglect the complex sociomateriality inherent in all creative endeavors. It overlooks not only the crucial interactions between creators and their artifacts but also the network of inter-objective relations (\citealp{latour_interobjectivity_1996})—the different tools, equipment, and media beyond the AI system—that collectively give rise to a creative product. While some frameworks do acknowledge materiality, they often treat it as an external factor. The “Five C's” framework, for instance, includes a context dimension, described as the “material surroundings of the work” but it is treated as external layers that influence the core collaboration rather than being part of the creative process itself (\citealp{kantosalo_five_2020}). The “rich surrounding for the co-creative collective, which thus interacts and affects the work of the collective in many ways” (\citealp[p.21]{kantosalo_five_2020}) and their diagram visually represents “context” as a containing shell that surrounds the interaction. This framing positions the material and social worlds as a stage upon which the cognitive process unfolds, not as an active participant in it.

\subsubsection{Interactions are limited only to the “AI-human dyad”}\label{Sec_2.1.2}

A more general consequence of these assumptions is the conceptual separation of interaction from context, limiting the scope of analysis to the human-AI dyad. The frameworks model interaction as a closed feedback between the human and the AI, focusing on their direct relationship and the shared artifact. This creates a sharp dichotomy between the “core” interaction and the “context,” detaching the co-creative system from the wider network of sociomaterial relations present in any creative practice. From an STS perspective, the boundary between an action and its background should not be treated as a pre-given, but as an outcome of practice that requires explanation. This act of framing is not merely conceptual but it is something materially done. A concert hall, for example, uses walls, doors, and acoustics to physically separate the performance inside from the urban space outside, thereby creating a localized site for a specific kind of interaction. As Latour (\citeyear{latour_reassembling_2007}) argues, it is precisely this work of materially assembling a frame that allows actors to give rise to a specific situation separating it from the world “outside.” 

Furthermore this limited dyadic focus is often described using idealized terms for interaction. High-level concepts like “the sharing of creative responsibility,” “synergistic creative process” or a "synergistic partnership" are frequently used (\citealp{kantosalo_five_2020}; \citealp{haase_human-ai_2024}; \citealp{davis_co-creative_2025}). While useful for high-level design, these terms reveal little about what actually happens when these systems are used. As STS research has shown, technologies are often reconfigured and reshaped by practices in ways that go beyond the uses anticipated by their developers (\citealp{hyysalo_method_2019}). Therefore, rather than taking these abstract descriptions of “collaboration” for granted, they should be investigated by focusing on diverse and situated creative practices.\footnote{It is important to note that frameworks such as COFI propose a more granular and nuanced vocabulary, breaking down concepts like “collaboration style” into specific components such as Participation Style (e.g., turn-taking vs. parallel) and Task Distribution (\citealp{rezwana_designing_2022}). However, even these more specific terms represent idealized modes of interaction. They pre-specify the possible modes of interactions, prescribing how interaction \textit{should} happen, rather than accounting for how it emerges and is reconfigured in actual practices.}

\subsubsection{The exceptionalism of AI agency}\label{Sec_2.1.3}

A third assumption concerns the exceptional status afforded to AI systems in comparison to previous technologies. The literature draws a sharp distinction between modern “active” AI systems and older “passive” digital tools. This framing posits GenAI as a unique technology possessing agency, a quality not typically ascribed to other artifacts. This is evident when Haase and Pokutta (\citeyear{haase_human-ai_2024}) state that while traditional Creativity Support Systems (CSS) “facilitate creativity,” they “remain tools rather than active contributors to the creative process.” Moruzzi and Margarido (\citeyear[p. 2]{moruzzi_user-centered_2024}) describe a paradigm shift that moves from “AI as a tool” toward “AI as an agent,” contrasting CSS where “the agency is in the hands of the human user” with computational creativity where “the artificial agent generally has more control.”

As has been noted by Bown (\citeyear{bown_beyond_2021}) and Moruzzi (\citeyear{moruzzi_creative_2022}) attributing creative agency solely to AI systems is misleading and “rather than asking whether an artificial actor can be ‘agentive’ or ‘creative,’ we should ask how agency and creativity are distributed in the network of actors who contribute to the process in question” (\citealp[p. 14]{moruzzi_creative_2022}). This is why it is better to talk about co-creativity, considering AI as “partners” or “collaborators” in the creative process. But these considerations should not overlook the broader understanding of technological agency as developed in STS. From this perspective, the capacity to act and shape outcomes is not exclusive to humans or AI systems. As ANT theorists argue, agency should be seen in the various artifacts that participate in practices, which are not mere passive tools for human intentions but active contributors and mediators to what actors do. The analysis should focus on the ability to “make a difference” in order to account for the different ways in which humans and non-humans contribute to the action (\citealp{sayes_actornetwork_2014}). As Latour argues we should ask: “Does it make a difference in the course of some other agent’s action or not? Is there some trial that allows someone to detect this difference?” (\citealp[p. 71]{latour_reassembling_2007}).\footnote{This, of course, leads to explaining action not only in terms of intentionality but also by considering all the entities that produce effects. Scholars like Schulz-Schaeffer (\citeyear{schulz-schaeffer_technology_2023}) have critiqued this minimalistic concept of agency for its limited heuristic power in distinguishing between actor types such as human and non-human ones. Latour’s minimal concept of agency is introduced here mainly as a way to question AI exceptionalism and to avoid analyzing action solely in terms of intentionality.}

Creating instead a rigid distinction between “active AI” and “passive tools,” the current discourse risks obscuring the significant contributions of other technologies. The goal should not be to erase the specific capabilities of GenAI, but to explain how its agency \textit{differs} from that of other technologies without creating an artificial divide. As we will show when discussing Filmmaking, most technological innovations—not only AI—are not simply tools supporting a director's vision, but actively shape the final product and its aesthetics.

\subsubsection{The prevailing approach is ontological but the attribution of creativity is multiple}\label{Sec_2.1.4}

A final and more fundamental assumption is the field's methodological reliance on an ontological approach to creativity. As Celis Bueno et al. (\citealp[p. 2]{celis_bueno_not_2024}) observe, these “approaches begin by asking ‘what is creativity,’ and once defined, proceed to evaluate whether a specific individual or machine fits under this definition.” Although initially formulated within the discourse on evaluating the creativity of autonomous AI systems, these considerations can be applied to the debate on co-creative systems. This is particularly evident in the field's reliance on cognitive definitions of creativity and its attempts to create fixed taxonomies for the possible interactions between humans and AI.

This ontological approach, however, often overlooks a crucial point: the central issue is not just what creativity is, but how it is ascribed and attributed. Creative agency is not an inherent property but is actively assigned to certain actors and actions through social, cultural, and technical processes. This “attributive agency” (\citealp{michael_actor-network_2017}) is established through credits (in publications, media, critics appreciation), legal devices (patents, copyrights), institutional rituals (awards, authorship order), and media representations (interviews, branding). For instance one can claim that a song is creative by the very fact that it is distributed by a label or that a writer can be regarded as particularly creative since praised by critics.

The intricacies of this process is especially clear in highly complex fields like the media industries. For decades, media industry studies have challenged the romantic view of the individual author, insisting that creative production is a collective effort shaped by numerous collaborators and industrial constraints (\citealp{redvall_authorship_2021}). The field highlights for example a stark division between “above-the-line” talent (directors, producers, writers), who are publicly framed as the primary authors, and “below-the-line” craftspeople and technicians, whose essential creative contributions are often invisible (\citealp{caldwell_authorship_2013}).

Understanding this long-standing complexity is also important to counter claims of AI exceptionalism concerning authorship and creative agency. The challenges of attributing authorship and distributing creative credit are not new problems introduced by GenAI. Rather, they are deeply entrenched, practical, and political issues central to most contemporary creative industries which are connected both to union policies, labor contracts and “negotiated interpersonally and collectively through a wide range of socio-professional rituals and habitual workaday routines” (\citealp[p. 350]{caldwell_authorship_2013}). Instead of treating GenAI as a unique case, we should analyze how it enters into these existing, complex networks of creativity attribution.

A second, overlapping issue concerns not \textit{who} is granted agency, but the valuation practices through which “creative” qualities are produced and attributed. These processes are not linear or single. Instead value is generated through practices that are complex, context-dependent, and can be acknowledged or contested by different actors (\citealp{hutter_three_2021}). These valuation practices are performed by a diverse field of “information intermediaries” who collectively shape what is deemed creative (\citealp{sharkey_expert_2023}) such as: expert critics and awards, who rely on professional knowledge and taste; ratings, rankings, and certifications, which use standardized criteria to create formal hierarchies; and online review aggregators, which draw on user-generated content.

The main point we want to make here is that the attribution of creativity and agency is a multiple and contested process. Ontological approaches, which attempt to establish a single, fixed definition of creativity, are ill-equipped to capture this dynamic multiplicity. This is not to say that particular accounts cannot become stabilized and transformed into standards or that agreements are never reached. Following STS, rather than accepting valuation practices as given, the focus should be on investigating their construction. Such an analysis may show how normative choices are made and how they determine who and what is included or excluded from the category of “creative."

\subsection{Distributed creativity between actors, spaces and times}\label{Sec_2.2}

Building on the critique of the assumptions that currently dominate the co-creativity debate, we will now outline a different approach for studying the role of GenAI in creative practices. As we have already shown, talking about AI and creativity cannot be limited to attributing creativity to a certain entity but requires focusing on the unfolding process going beyond mere cognitive approaches. Our aim is not to propose another general framework that provides definitions of creativity, AI and interactions. Instead, following STS approaches, we will provide a set of alternative assumptions and methodological orientations for the empirical investigation of creative production. The question is to make possible descriptions and analytical accounts of the process of creating something rather than giving a general framework of analysis (\citealp{sayes_actornetwork_2014}; \citealp{mattozzi_bodies_2021}).

Drawing on Celis Bueno et al. (\citeyear{celis_bueno_not_2024}), we propose a crucial shift in the research question: moving from an ontological inquiry of “what is creativity?” to an empirical investigation of “where and how is creativity practically accomplished?” This reframing leads directly to the concept of distributed creativity (\citealp{glaveanu_distributed_2014}). Rather than locating creativity in a particular actor (human or non-human), process (cognitive or otherwise), or a specific moment in space and time, the creative process is understood as something distributed and not traceable to a single locus. The precise nature of this distribution cannot be completely determined in advance. It must be empirically discovered by analyzing the specific creative practices and the network of human and non-human actors involved in any given situation. As suggested by Antoine Hennion an STS approach to creation should look at “what it means to produce things that did not exist. And to do so not by returning to metaphysical issues about creation, but by looking at the spaces, devices and techniques invented to allow invention. [...] as long as connections come before entities, 'creating' cannot mean producing something new from nobody knows where!” (\citealp[p. 73]{hennion_for_2015}).

To make a concrete case for this approach, we will use the example of Filmmaking. By analyzing key aspects of its production, specifically focusing on distributed agency and spatiotemporal dimensions of production, we will explain how distribution functions within creative practices.

\subsubsection{Agency is distributed between different actors}\label{Sec_2.2.1}

The first step in moving from co-creativity to a distributed perspective is to reconceptualize agency. A central way to conceptualize the agency of actors is to view them as mediators. Mediators are not passive intermediaries that simply transmit the artist's creative intention without affecting it (\citealp{latour_reassembling_2007}; \citealp{duff_assemblages_2017}). Instead, they are the active, constitutive elements of the creative process itself. The instruments, bodies, techniques, and objects involved "are neither mere carriers of the work, nor substitutes which dissolve its reality; they are the art itself" (\citealp[p. 89]{hennion_passion_2020}). This means that a creative outcome is the product of the entire network of mediators working together.

We can see this clearly in the practice of Filmmaking. A film's creativity is not located solely in the director's vision. It is distributed across a vast network of mediators both human (e.g., cinematographers, editors or "below-the-line" crew members) and non-human (e.g., the specific camera and lenses used, the editing software).\footnote{As this example illustrates, the distribution of agency does not imply an even or equal share of influence; rather, it is characterized by entrenched power asymmetries. Our focus remains on observing these attributions in action to reveal how agency is specifically negotiated and to whom it is assigned.} From this perspective, GenAI is not an exceptional "co-creator" but a new mediator entering this already complex network. Its agency lies in its capacity to "transform, translate, distort, and modify the meaning or the elements they are supposed to carry" (\citealp[p. 39]{latour_reassembling_2007}). Methodologically, to say that a non-human like an AI has agency is not to claim it has human-like intentions. By analyzing how GenAI actively reconfigures the creative process, we can move beyond AI vs other artifacts and begin to trace the distributed network that truly produces the work.

\subsubsection{Production is distributed across space and time}\label{Sec_2.2.2}

This focus on agency and practices immediately directs our analytical attention to the sites of production. Rather than locating creativity in the abstract cognitive processes of a human or the algorithms of an AI system, we must look at the "studio," what Farías and Wilkie (\citeyear{farias_studio_2015}) describe as the "humdrum and habitual workplace" where creation happens. The studio is not merely a physical space but a dense sociomaterial arrangement where practitioners "engage in conceiving, modelling, testing and finishing actual cultural artefacts."

Crucially, in complex creative fields like Filmmaking, the site of production is rarely a single, unified place. Instead, it is displaced and distributed across multiple, specialized spaces: pre-production offices where scripts are written and storyboards are drawn, various shooting locations, soundstages, and post-production houses where editing, sound design, and visual effects are completed. The studio therefore, is not a single container for creativity but is itself a distributed network of sites, each contributing to the process.

Creativity is also distributed temporally through a series of intermediary objects. Creative production rarely moves directly from an initial idea to a final product. Instead, it generates a series of temporary and provisional “objects” that allow the work to take shape over time. As Parolin and Pellegrinelli (\citealp[p. 438]{parolin_unpacking_2020}) note, creative actions produce "intermediary or temporary phenomena like sketches, diagrams and other models which help to think, prototype and trial aspects of what Beaubois calls the ‘object yet to arrive’." This concept is central to understanding production in the making. In the studio, creators work with what Hennion calls "maquettes"—sketches, drafts, and models that are "half an image made thing, half a thing made image" and serve as "an intermediary level of stabilization and resistance" (\citealp[p. 79]{hennion_for_2015}). These are not just representations of a final idea: they are active mediators that allow creators to test, negotiate, and refine the work as it develops.

This temporal distribution is clearly present in Filmmaking. A film's "final cut" appears as a stable, finished artifact but is the result of a long process involving countless intermediary objects such as multiple script drafts, daily rushes, rough cuts, and discarded scenes. Each of these temporary artifacts is a crucial site of creative work. Once the film is stabilized into a final cut, it achieves a new kind of autonomy, allowing it to circulate across different venues (cinemas, festivals, homes) and be reproduced through various devices (projectors, DVDs, streaming services).\footnote{It is important to keep in mind that this stabilization is relative. The identity of the film is reconfigured depending on the site of its reception. For example, in the context of a film festival it may be perceived as a "work of art." In a domestic setting it might be framed primarily as a form of "entertainment." The circulation of the artifact through these different venues is not a neutral process of distribution but an active process of mediation. Also the release of a restored version or a "director's cut" are all acts that materially reconfigure the film. These different versions challenge the notion of a single, definitive work and change how the film can be viewed and understood over time.}

A central site of distribution is found in what media production terms \textit{media assets}. These encompass the full range of intermediary materials—such as raw footage, sound recordings, and metadata—produced across diverse formats and sites of production. From a sociomaterial perspective, assets function as “intermediary objects” that circulate between professionals and production phases. Whether stored in physical archives or digital management systems, these assets constitute building blocks that are eventually synthesized during the editing process. As we will show in §\ref{Sec_4}, focusing on the assets is essential, as it is precisely at this level that the intervention of GenAI systems becomes most visible, transforming the audiovisual production.

\section{The role of technology in the aesthetics of Filmmaking}\label{Sec_3}

As we have shown in §\ref{Sec_2}, distributed creativity shifts the analytical focus from the finished artifact—the film—to the complex web of interactions between human agents and technological affordances that constitute the Filmmaking process. The focus is not only on who or what is acting but also on the actual spaces and temporalities that constitute the creative process.
In this section we trace some of the consequences of this approach for the study and analysis of the aesthetics of Filmmaking. As Mattozzi and Parolin (\citeyear{mattozzi_bodies_2021}) suggest, creative work could be understood from an STS perspective as “aesthetic practices” in which the act of making (\textit{poiesis}) and the resulting sensory experience (\textit{aisthesis}) are inextricably linked. From this perspective, the aesthetic qualities of a film—its style, form, and narrative texture—emerge not from isolated cognitive choices, but from the distributed practices and material conditions of its production.

To appreciate the implications of this shift, it is necessary to move beyond traditional film criticism and historiography, which often relegate technology to a mere functional background while overemphasizing individual authorship. Instead, this section argues that the specific configurations of distribution—how labor, tools, and environments are arranged—actively shape the cinematic style.

In §3.1, we examine how creativity is distributed across a multitude of professional roles, challenging the romanticized notion of the auteur as the sole source of creative output. Subsequently, §3.2 adopts a sociomaterial perspective to demonstrate how technological innovations have constantly redefined the practices of Filmmaking with direct effects on the aesthetics of cinema and on the very creative process. In other words, aesthetics is a consequence of the interaction among humans and technologies and its development can properly be appreciated only in terms of distribution.

\subsection{Beyond the Auteur: Filmmaking as a distributed process}\label{Sec_3.1}

Cinema is a collaborative art form that draws upon various artistic mediums, most notably photography, music, and theater. The process of live action Filmmaking is often highly complex and multifaceted, traditionally organized into three distinct stages: pre-production, production, and post-production. Pre-production concerns the different processes of screenwriting, funding, casting, location scouting, costume design, production design, etc. Production concerns the actual shooting process which involves principal photography and all the connected activities. Post-production concerns all the processes that come after principal photography, namely editing, scoring, dubbing, sound effects, sound design, color grading, subtitling, special effects, CGI, etc. Each of the aforementioned processes requires specific expertise. Also, the schematic division of the three main phases is purely theoretical: in practice, the various phases may overlap, occur at different times and influence each other through feedback processes.\footnote{An illustrative example can be found in television, where distribution requirements directly dictate narrative form. For decades, the structure of teleplays was governed by the temporal distribution of broadcast schedules. Screenwriters were compelled to organize scripts so that, every ten to twenty minutes, the story reached a climax or turning point. These transitions were not merely creative choices but functional necessities to allow for commercial breaks while maintaining audience engagement. Similarly, the use of \textit{cliffhangers} served to bridge the weekly gap between broadcasts. The advent of streaming services has since reshaped these conventions. Because audiences now engage in \textit{binge-watching}\textemdash where entire series are released and consumed at once—the need for frequent, scheduled climaxes has diminished.} Such feedback processes are also a consequence of technological innovations. For instance, commenting on the effects of digitalization, Caldwell (\citeyear{caldwell_authorship_2013}) talks about \textit{blurred and collapsed workflows}. By this, he refers to the way digital integration merges previously distinct stages of production—such as cinematography, visual effects, and editing—into a simultaneous and overlapping process. 

Despite this richness, film authorship is conventionally attributed to the director alone. While this attribution is both symbolic and functional, it is misleading when compared to the solitary authorship of a novelist. The director’s role is primarily one of coordination, translating a singular vision into a coherent result by managing a vast network of human and technological artifacts. In this sense, the director acts as an important decision-maker who selects from an array of aesthetic and technical options provided by various collaborators. Attributing authorship to the director is appropriate only insofar as it identifies the individual entrusted with the final “vision” and the mastery of cinematic language, yet it obscures the distributed labor that constitutes the actual Filmmaking process.

A central notion in film criticism is the notion of \textit{auteur}, which is French for \textit{author}. In the Golden Age of Hollywood cinema (about the 1930s to the early 1960s), the director was regarded as a technical figure rather than a creative one. It was the work of French critics between the late 1950s and early 1960s that changed such an attitude. Specifically, critics associated with \textit{Cahiers du cinéma} sought to demonstrate that certain directors were able to express a distinct personality and vision through technical and aesthetic choices. This was in line with the concept of the \textit{caméra-stylo} introduced by French critic Alexandre Astruc (1948), which metaphorically equated the camera to a pen and the director to a writer. A first clear description of the new approach to film criticism was provided by François Truffaut in an article titled \textit{Une certaine tendance du cinéma français} (\citeyear{Truffaut_tendance_1954}). The next year, in a review of the film \textit{Ali Baba and the Forty Thieves} [\textit{Ali Baba et les 40 voleurs}, 1954] directed by Jacques Becker, Truffaut (\citeyear{Truffaut_Baba_politique_1955}) introduced the expression of \textit{politiques des auteurs} (French for \textit{politics of authors}). In 1962, the American film critic Andrew Sarris introduced the notion of the auteur to the English-speaking world with his essay \textit{Notes on the Auteur Theory} (\citeyear{Sarris_auteur_theory_1962}). Indeed, the preference for the term “theory” over “politics” resulted in a misleading translation that led to a significant semantic shift. The main objective of the \textit{Cahiers} critics was not to formulate a theory of authorship, nor to demonstrate that the director is the sole author or that every film serves as an expression of individual genius and creativity. As Jean-Luc Godard and others frequently observed, the \textit{politique des auteurs} aimed to recognize the distinct artistic personality of directors like Alfred Hitchcock or Howard Hawks. At the time, these figures were often dismissed as mere makers of “B-movies” or commercial entertainment, supposedly incapable of engaging with profound philosophical or political themes.

As pointed out by Peter Wollen (\citeyear{wollen_signs_2013}), Sarris’ work eventually led to many misunderstandings about the notion of \textit{auteur} since the auteur theory was never elaborated in programmatic terms. Some critics went even further distinguishing a proper \textit{auteur}—a director whose work holds a semantic dimension—from a \textit{metteur en scène}—a director whose focus is essentially on style. In his analysis, Wollen shows how in the end the auteur theory is essentially a lens for critical interpretation since it “implies an operation of decipherment; it reveals an author where none had been seen before” (\citealp[p. 61]{wollen_signs_2013}). 

From an STS perspective, attributing the authorship of a film to the director can be understood as a form of boundary work. As defined by Gieryn (\citeyear{gieryn_boundary-work_1983}), this practice involves tracing boundaries between different domains of labor to assign specific responsibilities and legal rights. This designation of the director as the “author” leads to significant practical implications. On the one hand, it provides film critics with a framework for interpreting meaning consistent with the auteur theory—a discourse that relies on the established existence of an author for its own credibility. On the other hand, this conceptualization serves the strategic interests of both directors and producers. In fact, the director’s prestige and authority are rooted in practical, commercial, and institutional dynamics. For the sake of production fluidity, it is often said that the director must be granted the status of a “dictator” on set: their choices are to be debated but not questioned. However the degree to which an individual possesses control to shape a production is heavily dependent on the industrial context and the scale of production. While the director might exercise command on the set, the ultimate agency, represented by the “final cut,” varies significantly: in the US studio system, this executive power is often held by the studio or producer, whereas in many European cultures, the director is symbolically regarded as the primary author and has more agency on the production (\citealp{redvall_authorship_2021}).

Furthermore, the narrative of the director as the author of the film sustains a rigid division between below-the-line (BTL) and above-the-line (ATL) workers (\citealp{caldwell_authorship_2013}). Such expressions refer to the habits of Hollywood productions, and distinguish workers who work on the physical production at fixed hourly rates from creative talents and management, such as screenwriters or directors. Such distinction is certainly rigid and scholars argue that it is not suitable for a proper analysis of film working practices (\citealp{antoniazzi_media_2024}), nonetheless it is helpful to start a discussion on the attribution of merit and prestige. While the director’s central role in Filmmaking is undeniable, it is equally necessary to recognize that every participant in the process potentially exerts significant influence. Nevertheless, BTL workers are frequently stripped of their creative agency in public discourse (\citealp{caldwell_authorship_2013}). Following an STS perspective, creativity can be seen as a distributed and negotiated process involving all participants across every phase of Filmmaking. However, this process is not limited to human interactions alone. As §\ref{Sec_3.2} will show, Filmmaking is also deeply shaped by the various technologies utilized during both shooting and editing. It is precisely within these sociomaterial interactions that the act of directing unfolds. An analysis of these interplays reveals that directing is not a monolithic practice, but rather a process encompassing diverse approaches.

\subsection{The aesthetic implications of technology on Filmmaking}\label{Sec_3.2}

Ultimately, Filmmaking is the process of producing and assembling images and sounds into a cohesive whole. Consequently, a film's aesthetic is the cumulative result of every decision made during production: from a specific camera movement, the timing of a cut or a specific line of dialogue. Each of these choices constitutes a distinct creative act. Therefore, a proper in-depth analysis of film cannot avoid investigating the techniques and the interactions involved in the process, and showing how they relate to style and aesthetics. 

Relevant works that show the connections between technology, technical choices and aesthetics are those of Barry Salt, Christopher Beach and David Bordwell. Salt draws a connection between the developments in film technology and the development of the formal features of film style (\citeyear{salt_film_2009}). Such connection is strongly supported by a thorough historiography of film technologies and statistical analyses of their uses.\footnote{Salt’s statistical analysis of film (\citeyear{salt_statistical_1974}) adopts a radical approach in taking notes of such elements as the length of the shots, the camera movements, the framing of the shots, etc.} Salt's statistics on film style are organized mainly by decade and focus mostly on mainstream Hollywood cinema; nonetheless they are sufficient to prove his point and one could easily extend them to a global analysis and to all forms of cinema and audio-visual content.
Beach instead stresses how cinema is a collaborative practice and focuses his analyses on the peculiar collaboration between the director and the director of photography (DOP). Such an approach is of course limiting in an STS perspective which stresses the importance of considering all the actors involved. However, it is significant and relevant for the present discussion. Beach proves that “in the creation of cinema, technological development and aesthetic elements build upon and mediate each other through the process of collaboration” (\citealp[p. 2]{beach_hidden_2015}). In a chapter about the collaborations of DOP Gregg Toland with directors William Wyler and Orson Welles, Beach demonstrates Toland’s effort in developing new techniques to perfect deep-focus cinematography allowing for new staging and composition possibilities until then impossible. 
As much as celebrating Toland’s genius and initiative, Beach stresses that in the studio era of Hollywood production cinematographers were typically given specific indications about which lighting techniques to adopt in order to obtain a recognizable visual style as a trademark of the production companies themselves. Eventually, the advent of a new generation of cinematographers\footnote{Beach’s work is further supported by many in-depth interviews with cinematographers by Schaefer and Salvato (\citeyear{schaefer2013masters}).} in the 1960s and 1970s who embraced new techniques and camera equipment led to new changes in Hollywood visual styles showing the interactions between technology and aesthetics. 
Finally, Bordwell deconstructs many usual approaches to film history and focuses his attention on film style stating that he takes style “to be a film’s systematic and significant use of techniques of the medium” (\citeyear[p. 4]{bordwell_history_2018}). In an extensive chapter Bordwell shows how “the history of depth staging intersects with histories of technology [...] and of production practices” (p.269).

As these studies demonstrate, technologies have been deployed in diverse ways to achieve specific aesthetic outcomes. The choices taken in Filmmaking may be either guided by stylistic choices\footnote{To appreciate the stylistic choices that guide the practices of Filmmaking, one can take a look at the manuals adopted in film schools about directing (\citealp{katz1991film}; \citealp{rabiger_directing_2013}), cinematography (\citealp{alton_painting_1995}; \citealp{brown_motion_2008}; \citealp{brown_cinematography_2012}; \citealp{brown_filmmakers_2015}) and editing (\citealp{reisz2017technique}).} (the explicit desire to obtain certain aesthetics effects) or by practical constraints\footnote{The implications of practical constraints is well-documented in the firsthand accounts of filmmakers reflecting on their production experiences. Prominent examples include the published production diaries of Spike Lee for \textit{She’s Gotta Have It} (\citeyear{lee_spike_1987}), John Sayles for \textit{Matewan} (\citeyear{sayles_thinking_1987}), and Robert Rodriguez’s account of \textit{El Mariachi} (\citeyear{rodriguez_rebel_1996}). Similarly, Emma Thompson’s (\citeyear{thompson_sense_1996}) diary regarding \textit{Sense and Sensibility} provides critical insight into the intersection of screenwriting and performance under the direction of Ang Lee. Beyond diaries, foundational Filmmaking manuals by directors such as Sidney Lumet (\citeyear{lumet_making_1995}) and Alexander Mackendrick (\citeyear{mackendrick_film-making_2005}) offer invaluable perspectives. Both Lumet and Mackendrick draw extensively on their professional careers to testify to the constant negotiation between theoretical vision and the pragmatic demands of the film set.} (e.g., the pressure of concluding a scene before sunset or the logistical limitations imposed by available camera equipment). Furthermore, some of the most celebrated creative choices arise not from rigid planning, but from the capacity to improvise on the spot, as creativity can arise within the specific configurations that occur in the production process.\footnote{This is the case for many iconic lines, as in the scene of Robert De Niro talking to himself in the mirror in \textit{Taxi Driver} [1976], or in the scene of Rutger Hauer during the final showdown with Harrison Ford in \textit{Blade Runner} [1982].} How these technological and practical aspects are navigated depends also on the managerial and artistic approach of the director who might choose to employ a specific style defined by rigorous technical choices.\footnote{For instance, Japanese director Yasujirō Ozu eventually developed a very rigorous style shooting only fixed shots with a 50mm lens; he was very careful about framing, paid no attention to the 180 degree rule and had his actors perform each take many many times (\citealp{ozu_scritti_2016}; \citealp{yoshida_ozus_2003}). The 180 degree rule states that when shooting conversations between two characters, shots should be always taken from only one side of the imaginary line connecting the eyes of the two characters. Otherwise, in the editing process it would appear that the characters looked not at each other but in different directions which could be confusing. Actually, rather than a rule, it is a principle: more a rule of thumb than a law; juxtaposition with proper establishing shots or the use of appropriate camera movement can avoid any sense of confusion. Nonetheless, the work of Ozu stands out since his choices in framing, although effective, are quite counterintuitive.} There are many types of directors: some have a deep understanding of any technical aspect and are very demanding to their collaborators.\footnote{This is exemplified by directors such as Stanley Kubrick (\citealp{castle_stanley_2016}).} Some are more judges to the work of their collaborators and just guide and overview the Filmmaking process.\footnote{This is characteristic of Woody Allen, who has repeatedly emphasized his dependence on frequent collaborators and his practice of allowing actors substantial room for interpretation (\citealp{lax_conversations_2007}; \citealp{lax_start_2017}).} But, in line with STS, one should not consider film technologies as simple tools in the hands of professionals but as mediators which both afford and constrain Filmmaking practices, and consequently produce specific aesthetics results (\citealp{latour_reassembling_2007}; \citealp{hennion_passion_2020}). In §\ref{Sec_3.2.1}, we will present some examples from the history of film technology in order to make this point clear.

\subsubsection{Relevant examples from past technological innovations}\label{Sec_3.2.1}

To illustrate the intricate relationship between technology and aesthetics, this section examines three relevant examples: the evolution of sound recording, the development of camera movement equipment, and the use of \textit{performance capture} in digital production. While film history often highlights the introduction of sound, the introduction of color, and the digital turn as the medium’s primary leaps, Peter Wollen (\citeyear{wollen_cinema_1980}) offered an acute analysis challenging this perspective. He argued that such a simplified view of technological development is a common limitation of film criticism. In reality, many innovations with a profound impact on film style often go unnoticed such as the introduction of adhesive tape (Scotch tape) in the editing room, which fundamentally altered the physical and creative process of cutting film. 
The examples provided here are intended to be illustrative rather than exhaustive and many other innovations could be cited. The main purpose is to highlight the sociomaterial dynamics inherent in film production and to pose broader questions regarding the co-evolution of technology and aesthetics. Establishing this historical framework is essential for accurately situating GenAI systems and investigating the implications of these emerging innovations.

Let’s first consider the process of recording sound on set. To prevent unwanted noise, the production environment requires strict discipline; during a take, everyone except the actors must remain completely silent. Furthermore, the placement of microphones is often a major challenge that does more than just capture audio: it can actually dictate the visual framing of a shot or restrict how the actors move. The first microphones employed in cinema were big enough to require them to be hidden by other objects on set. In many early sound films, dialogue scenes set at tables often featured large flower vases used specifically to conceal microphones. As a result, actors were forced to lean toward these vases while delivering their lines and this rigid style of movement remains visible when watching Hollywood films from the 1930s.
Eventually, directional microphones would be attached to boom poles that are held by a proper boom operator on set during shootings. The role of boom operators significantly reshaped the practices on set. Camera movements need to be carefully choreographed not only to follow the actors but also not to get the boom within the frame. As Beach (\citeyear{beach_hidden_2015}) observed, the introduction of sound produced several unintended consequences for Hollywood’s visual style. Because early cameras were noisy, they had to be isolated within soundproof booths, which effectively eliminated camera movement and restricted scenes to fixed shots. Furthermore, cinematographers were forced to use long focal-length lenses, which significantly reduced the depth of field. Even lighting equipment posed a challenge; the noise generated by certain lights necessitated the adoption of entirely new lighting techniques and technologies.

In addition to sound, the development of equipment for camera movement represents a second major domain where technology has directly reshaped film aesthetics. Small and light cameras can be moved handheld by means of camera grips or shoulder rigs. Stable movements can be obtained with proper camera gimbals or steadicams. The effect on the shots is different and provides them with peculiar feelings. A shot obtained with a steadicam is different from one obtained with a dolly. Other than the framing, it is the very movement to be different even if they are both stable. Yet, the steadicam requires just one operator to move the camera—forgetting about the focus puller—while dollies need the coordination of many operators. Even more, handheld movements may require cheaper equipment than stabilized movements. Therefore, in the end handheld shots are cheaper and faster to realize and many times they are adopted over more complicated shots just to save time and money and not for aesthetics reasons (\citealp{salt_film_2009}). Nonetheless, these innovations significantly have effects on film aesthetics by reshaping established production practices. This transition can also result in a decline of traditional craftsmanship. As new methods become standardized, the high level expertise needed for the old techniques can get lost.

A final and recent example of the aesthetic possibilities consequent to technological development is provided by the use of performance capture techniques in the making of the Avatar franchise films directed by James Cameron, namely \textit{Avatar} (2009), \textit{Avatar: The Way of Water} (2022) and \textit{Avatar: Fire and Ash} (2025). \textit{Performance capture} is an expression coined to describe an extension of \textit{motion capture}. \textit{Motion capture} consists in the recording of the movements of an actor or of an object by means of sensors attached to it. \textit{Performance capture} refers to a complete recording of the movements of the body, the facial expressions and the voice of an actor moving in a three-dimensional space, therefore a digital recording of the entire performance of an actor. The technology developed by Cameron’s team makes it possible to then use the recorded data to reconstruct a digital version of the performance of the actor, substitute their body with a digital one and place it in a digital environment with a background entirely created in post-production.\footnote{See for instance the following interviews to James Cameron on YouTube: \href{https://www.youtube.com/watch?v=tHDs-4quso0}{https://www.youtube.com/watch?v=tHDs-4quso0}; \href{https://www.youtube.com/watch?v=qSbBg56k4HA}{https://www.youtube.com/watch?v=qSbBg56k4HA}.} Most significantly, once the actors' performances are digitally captured, the camera angles and shot compositions can be determined entirely during post-production. This shift eliminates the traditional need to spend hours on set positioning cameras, rehearsing complex movements, or adjusting lenses and lighting. As a result, actors are granted a new level of creative freedom, allowing them to remain fully immersed in their performance without the technical interruptions of a conventional film set. \textit{Performance capture} enables a complete separation of the traditional phases of \textit{blocking} (i.e., the choreography of the movements of the actors on set) and \textit{staging} (i.e., the positioning of the camera on set, therefore the decision of the shots and the realization of the découpage) leading to new aesthetic possibilities.

\subsection{Disruption or continuity? From past technologies to AI techniques}\label{Sec_3.3}

The previous discussion shows that film aesthetics is a consequence of all the choices taken in the Filmmaking process, choices that require the expertise of various professional figures capable of handling the technologies used in the process itself. We have seen how the introduction of new technologies has constantly reshaped Filmmaking practices with consequent effects on the aesthetics. We have emphasized how affordances simultaneously provide new creative opportunities and impose practical constraints on Filmmaking. 

It can be argued that AI-driven techniques represent a new technological frontier that will yield comparable effects. However, at present, there is a lack of sufficient empirical research to fully assess the consequences of GenAI systems on production. Such research has already been carried out for the effects of other digital technologies. For instance, Swords and Willment (\citeyear{swords_it_2024}) have investigated how virtual production has challenged traditional production models, and allowed post-production possibilities to dictate pre-production and production working practices. It is reasonable to expect similar consequences with AI. Nonetheless, it remains difficult to discern whether AI represents a continuity of the trajectory of digitalization or if it constitutes a disruptive phenomenon that will radically restructure production (\citealp{erickson_ai_2024}). Among the dynamics already established by digitalization are the increasing precarity of creative labor and a lowering of entry barriers that intensifies market competition. Whether AI will further exacerbate these trends by fully replacing workers or merely displacing them within the production chain remains an open question. Beyond these labor concerns, it is even more complex to understand how these new production dynamics will reshape the aesthetic possibilities of the medium.
Nonetheless, on the basis of the previous discussion, we can introduce a proper lens of analysis and make some remarks on the effects of specific AI techniques. The elements we outlined in the discussion on previous technologies can now be formulated as open questions suitable as categories for the critical analysis on the effects of AI on Filmmaking we will develop in §\ref{Sec_4}. Such questions are:

\begin{itemize}
    \item How are Filmmaking practices reconfigured by the introduction of AI techniques?
    \item How does the aesthetics of film change as a consequence of the introduction of AI techniques?
\end{itemize}

\section{The integration of AI techniques in Filmmaking} \label{Sec_4}

Current discourse on GenAI in Filmmaking primarily focuses on comparing the capabilities of text-to-image and text-to-video models with traditional live-action production. This comparison has fueled preoccupations over the potential replacement of workers across production workflows. These perspectives, often based on a deterministic understanding of GenAI, view technological artifacts as having inevitable and unavoidable effects for working practices. However, this viewpoint overlooks how technologies are integrated into daily professional interactions, reconfiguring “who crosses paths with whom, when, and for what purpose” (\citealp[p. 4]{baygi_generative_2025}). 
These fears also conflict with the current technical possibilities, since the models still struggle to maintain both spatial and temporal consistency. For instance, advanced models like Sora often struggle with the precise arrangement of elements (sometimes confusing basic directions like left and right) or fail to maintain the intended temporal flow of a scene during complex camera movements. The system may even spontaneously insert irrelevant characters or objects, altering the planned composition and atmosphere of the narrative (\citealp{liu_sora_2024}). Furthermore, current systems lack the necessary fine-grained control for professional use, necessitating extensive manual refinements and multiple generative iterations to mitigate existing technical limitations (\citealp{zhang_generative_2025}).

Building on the preceding considerations, this section argues that the effects of AI on Filmmaking are better understood by examining how specific techniques are integrated into and reconfigure situated practices. Rather than proposing far-reaching speculative scenarios, this analysis explores the potential consequences of GenAI by grounding the discussion in the reflections on distributed creativity, working practices, and aesthetic effects established in previous sections. We hold that the best way to advance our argument is to focus on the manipulation and generation of media assets (i.e., images, sounds and videos). As a matter of fact, these assets are intermediary objects (see §\ref{Sec_2.2}): they are the building blocks ultimately to be assembled in order to obtain the film and their production is the consequence of the interactions among diverse professional figures. As argued in §\ref{Sec_3}, new technological developments reconfigure the production pipeline as well as the aesthetics of the final work. Consequently, showing how specific AI techniques transform assets allows us to appreciate the resulting shifts in practice.

To categorize these effects, we propose a tentative taxonomy that distinguishes different categories of AI techniques and highlights their influence on production. Our use of the term “technique” extends beyond the GenAI model or system itself. We define it as the application of technology to specific practical purposes. Put simply, a technique describes how GenAI systems can be used to manipulate and generate media assets within the production process. We identify three primary families of AI techniques\footnote{Please note that the same AI model or system can be used to implement techniques that belong to different families.} based on their role in the asset production workflow: \textit{asset enhancement}, \textit{asset editing}, and \textit{asset generation}. Asset generation techniques, such as text-to-video synthesis of entire scenes, stand out from asset enhancement and asset editing techniques in that they gained the most attention both from the industry and media coverage. As will be illustrated, both asset enhancement and asset editing techniques still work on the same principle: taking a real production asset and enhancing or augmenting some of its properties to better suit a production or artistic need. While they offer new approaches to Filmmaking, asset enhancement and editing remain the least utilized and known generative techniques. Outside of major productions with significant VFX budgets, their adoption is likely limited by the high level of technical complexity and specialized know-how required for a correct implementation. On the other hand, much of the attention nowadays is dedicated to asset generation systems, which have rapidly entered the toolkit of pretty much every filmmaker and studio in the world. Such rapid diffusion might be the consequence of two factors: the ease of use of these generative models given by the commercial interfaces that have been developed, and the intrinsic paradigm shift in media generation that they bring to the table where a piece of media now can be produced with just a string of text.

The following sections detail each family by examining:

\begin{itemize}
    \item The features of the different techniques and some uses in production;
    \item The ways in which their integration could reconfigure established working practices;
    \item The potential aesthetic consequences these techniques have on the final film.
\end{itemize}

These dynamics will be summarized in three tables providing for each family a synthesis of technical descriptions, reconfigurations of practices, and potential aesthetic effects. The tables are presented in the appendix, namely table \ref{Tab_1} for asset enhancement techniques, table \ref{Tab_2} for asset editing techniques and table \ref{Tab_3} for asset generation techniques.

\subsection{Asset enhancement techniques}\label{Sec_4.1}

We classify as asset enhancement techniques those techniques that do not replace traditional components of the production pipeline but instead augment existing audio-visual materials, extracting additional value from them. Rather than supplanting creative or technical roles, asset enhancement techniques enhance the usability, quality, and adaptability of pre-existing content, facilitating higher production value with lower cost or labor intensity. Examples of asset enhancement techniques are \textit{video upscaling}, \textit{frame interpolation}, \textit{audio separation}, \textit{audio denoising} and \textit{normal} and \textit{depth maps estimation}.

\subsubsection{Technological features and uses}\label{Sec_4.1.1}

\textit{Video upscaling} refers to the use of AI-driven super-resolution models that increase the resolution of video footage, often transforming standard definition or high definition content into ultra-high definition formats such as 4K or 8K. These models learn to hallucinate (i.e., statistically infer) plausible high-frequency details absent from the original footage, allowing archival or low-budget content to meet contemporary visual standards. Upscaling is particularly valuable in remastering older films or adapting assets for new distribution platforms (\citealp{wang_edvr_2019}).

\textit{Frame interpolation} refers to the generation of intermediate frames between existing ones by AI models. This is typically used to convert lower frame rate footage (e.g., 24fps or 30fps) into smoother high frame rate versions, such as 60fps or even 120fps (\citealp{huang_real-time_2022}). Yet, one could also apply frame interpolation models to footage shot at relatively high frame rates (e.g., 60 fps) and produce ultra-slow-motion sequences with unprecedented temporal resolution. This kind of hyper-slow motion, generated synthetically rather than captured through expensive high-speed cameras, could offer new aesthetic possibilities for storytelling, sports analysis, scientific visualization, and more. Frame interpolation enhances the visual fluidity of motion and is especially useful in action sequences, animation, and virtual reality content. These interpolative models rely on motion estimation algorithms that predict object trajectories and occlusions between frames, resulting in more natural movement and improved viewer experience. Again, frame interpolation plays a key role in film restoration through its ability to regenerate damaged frames (\citealp{andreose_plausibility_2025}).

\textit{Audio separation} refers to the use of models designed to isolate individual sound sources from a mixed audio track. For instance, such models can separate dialogue, music, ambient noise, or sound effects from a single audio stream. Also, they support more flexible localization (e.g., dubbing or subtitling) and adaptive sound design workflows. These capabilities are instrumental in re-mixing or restoring older or independent films, where access to isolated stems is often unavailable. In fact, audio separation can isolate speech from noisy historical recordings, enabling clear subtitling, translation, or even voice cloning for restoration or narration.

\textit{Audio denoising}, on the other hand, refers to the use of models aimed to suppress unwanted background noise or artifacts while preserving the integrity of the target signal, typically dialogue. Trained on extensive datasets of clean and noisy recordings, these models can significantly improve audio clarity in recordings affected by environmental interference, low-quality microphones, or compression artifacts (e.g., \citealp{schroter_deepfilternet_2022}). In Filmmaking, this allows for greater salvageability of audio production, reducing the need for ADR (Automated Dialogue Replacement) sessions and streamlining the post-production process. Also, when applied to a track obtained with audio separation techniques, denoising further enhances intelligibility without the artifacts common in traditional filters.

\textit{Normal} and \textit{depth maps estimation} refer to the use of \textit{normal} and \textit{depth maps estimators}, a peculiar type of models that predict additional information from 2D video data (e.g., depth maps contain distances between objects and the camera) resulting in novel post-production possibilities (e.g., \citealp{yang_depth_2024}). Through these kinds of predicted inputs, standard 2D video can be better color corrected/graded and manipulated to feature entirely new lighting scenarios and lens emulation, thus greatly improving on the strict technical and economical constraints of traditional film equipment.

\subsubsection{Sociomaterial aspects}\label{Sec_4.1.2}

\textit{Equipment quality}. Asset enhancement techniques can augment the quality of the assets produced in the shooting and allow for finer manipulation in the editing process. On the one hand, this enables the production of high quality products even when shooting with low quality equipment. On the other hand, new editing possibilities will lead to a different approach to cinematography and directing. By enabling the use of older or lower-end recording equipment without compromising final output quality, asset editing techniques could make it easier to access professional-grade production tools. This shift will lower the economic and technical barriers to entry in Filmmaking and likely allow independent filmmakers, low-budget productions, and archival projects to achieve results that were previously exclusive to high-end studios with access to state-of-the-art resources.

\textit{Film restoration}. The application of asset enhancement techniques to the restoration process is significant. Upscaling models can transform historically significant, low-resolution footage into visually coherent material suitable for contemporary screens. Combined with audio separation and enhancement techniques, this makes it possible to repurpose archival content, such as early 20th-century film reels, ethnographic recordings, or degraded newsreels, in modern cinematic or documentary contexts. These capabilities not only preserve cultural heritage but also provide raw material for new narratives. Filmmakers can interweave restored archival elements into modern productions, blend historical voices with contemporary sound design, or use old footage to create speculative or anachronistic aesthetics. Thus, AI asset enhancement models serve both a restorative and generative function, reviving the past while simultaneously expanding the creative vocabulary of the present.

\textit{Découpage control}. Other than improving image quality, these techniques will redistribute creative agency across the production network. By allowing for digital recomposition and enhancement in post-production, these systems reduce the necessity for technical perfection on set, effectively distributing the locus of aesthetic control between the camera and the edit suite. For instance, the possibility to select a part of a shot and enhance its quality thanks to video upscaling could lead people to pay less attention to composition on set just to better compose zooming in during the editing. Also, it will be possible to shoot a medium shot and cut a close-up from it. This could lead to the choice of not using long lenses for close-ups, making it possible to leave more room for the actors to move freely when they are performing for a close-up. Furthermore, filmmakers can adjust shot dimensions without swapping lenses, saving significant time on set by reducing the need for lens changes and lighting recalibration. However, the physical perspective of the final shot will differ from one obtained through traditional lens changes, ultimately resulting in a distinct aesthetic.

\textit{Frame rate control}. Since frame interpolation will make it possible to obtain higher frame rates videos from those shots on set, it will be possible to make convenient choices during production. One will be able to shoot at a lower frame rate and maintain a higher time of exposure making it possible to shoot the same scene with less light; for instance if one shoots at 24fps instead of 48fps the cinematographer can save up to a stop of light. Furthermore, shooting at a lower frame rate reduces the volume of recorded data or shooted film, potentially streamlining the editing process. This approach allows filmmakers to defer decisions regarding slow-motion sequences until post-production, selecting specific moments for temporal manipulation after the shoot. While this may be limiting because filmmakers miss seeing slow-motion dailies during the shoot, it is also freeing, as slow motion can be applied even to footage recorded at low frame rates.

\textit{Sound manipulation}. Many times shootings are limited by the presence of undesired noises. Also, good takes can be ruined by a single noise that would force the entire crew to shoot another take. Techniques such as audio separation or audio denoising will allow shootings in noisy environments and prevent the need for extra takes.

More generally, these possibilities will enhance a collapse of the workflows: editing will influence ever more directing and cinematography as it already happened with the advent of digital (one could think for instance about the influence of the color grading process on lighting and on the use of filters). Table \ref{Tab_1} provides a synthesis of technical descriptions, reconfigurations of practices, and potential aesthetic effects for asset enhancement techniques.
\vspace{0.5cm}

\subsection{Asset editing techniques}\label{Sec_4.2}

We classify as asset editing techniques those systems that use GenAI models for advanced editing tasks such as: modify, rearrange, or replace specific elements within existing audiovisual materials. These tasks require a manipulation of photographed images and recorded sounds that does not extract additional value from the already existing audio-visual materials but builds on them. Examples of asset editing techniques are \textit{in-painting}, \textit{out-painting} and \textit{face transfer}.

\subsubsection{Technological features and uses}\label{Sec_4.2.1}

\textit{In-painting} refers to the deployment of image generation models for the filling in of missing or occluded areas of an image (\citealp{po_state_2024}). On the other hand, \textit{out-painting} refers to the use of such models to extend the boundaries of an image beyond its original frame. These capabilities have powerful implications for Filmmaking. In-painting enables the seamless removal or replacement of unwanted objects, characters, or artifacts from individual frames or entire sequences, facilitating continuity correction or post-production cleanup without the need for re-shoots or extensive manual rotoscoping. Out-painting, conversely, allows post-production specialists to expand a scene spatially, for example by synthetically generating backgrounds, architecture, or landscape features beyond the original camera field. This can be particularly useful in virtual production, VFX-heavy sequences, or the adaptation of footage to different aspect ratios. More broadly, these models allow for entirely new approaches to scene construction. For example, generative image models can synthesize entire environments, props, or even character likenesses, which can then be composited into live action or animated sequences. This reduces dependency on physical sets or CGI workflows (\citealp{singh_role_2025}), especially in previsualization or rapid prototyping contexts.

\textit{Face transfer} refers to the use of models that enable the precise transfer of facial expressions, movements, or entire identities from one actor to another in video sequences (\citealp{tolosana_deepfakes_2020}). These models can map performance data from a source actor onto a target face while preserving emotional nuance and realism. In Filmmaking, this technology offers a range of positive applications for actors. For instance, it can allow stunt doubles to perform dangerous sequences while seamlessly retaining the lead actor’s likeness, thereby enhancing safety without compromising narrative continuity. It also opens up new possibilities for localization, enabling actors' faces to be re-synced to match dubbed dialogue more naturally across languages. Additionally, face transfer can extend an actor’s presence in a project, such as enabling consistent performance across reshoots or aging/de-aging characters without extensive prosthetics or makeup. Far from replacing actors, these tools can actually enhance their performances and preserve them in increasingly adaptive and cross-platform storytelling environments. Also, they could enable more flexible scheduling.

\subsubsection{Sociomaterial aspects}\label{Sec_4.2.2}

\textit{Shot composition}. In-painting and out-painting allow editing considerations to influence set design early on. Because unwanted elements can be digitally removed or modified, filmmakers can shoot in authentic scenarios with significantly fewer physical interventions on set. For instance, for outdoor shootings one will not have to remove or cover signs, car plates or passersby which are typically annoying actions. This should free cinematographers and set designers by reducing the time and the materials needed to prepare the set. Also, independent filmmakers will be able to shoot outdoors without the need to comply with the bureaucracy of permits. Knowing that the background of a shot can be extended will make it possible to obtain wide shots from medium or close shots. For instance, one could be shooting on a sound stage with the actor against a fake wall and then extend the shot to show the actor in front of a house and a larger background (a field, other houses, a street); consequently set designers could build smaller sets.

\textit{Face transfer}. The application of face transfer technology will likely facilitate the integration of stunt doubles into shots, liberating cinematographers and makeup artists from traditional constraints. By removing the need to physically disguise doubles, these tools allow for greater creative freedom in shot composition while ensuring that the transition between the stunt performer and the lead actor remains seamless to the audience. Also, it will be easier to shoot scenes in which an actor had to play twins: one could now just hire a second actor for the role of the twin and easily superimpose the face of the main actor. Furthermore, with in-painting and out-painting one can easily solve continuity mistakes by manipulating single frames and avoiding the need for re-shoots.

In light of these developments, we may witness a merging of workflows where professional roles that previously interacted with the director in isolation begin to collaborate directly. The effects of asset editing techniques are first seen in post-production, yet pre-production and shooting could adapt to integrate these capabilities into the workflow. This shift allows cinematographers, editors, and designers to propose collective creative choices, fostering a non-linear approach to production before the shooting. Table \ref{Tab_2} provides a synthesis of technical descriptions, reconfigurations of practices, and potential aesthetic effects for asset editing techniques.

\subsection{Asset generation techniques}\label{Sec_4.3}

We classify as asset generation techniques those techniques that create content ex nihilo via direct request. Typically, requests are made through textual prompts and the models generate texts, images, videos or sounds. Examples are generative text-to-image models like Stable Diffusion XL and Flux.1 or text-to-video models such as OpenAI Sora and DeepMind’s Veo3.

\subsubsection{Technological features and uses}\label{Sec_4.3.1}

While asset generation is the preferred use of generative models, they allow for many different input modalities, which help creators to move from abstract conceptualization to precise aesthetic control. These modalities determine the degree of predictability and granular influence a user has over the final output.

\textit{Text-to-Image} (T2I). In this modality GenAI models translate natural language prompts into visual assets. While powerful for rapid ideation and concept art, T2I often suffers from prompt engineering limitations, where the stochastic nature of the model makes it difficult to achieve exact spatial layouts or specific character consistency through text alone.

\textit{Image-to-Image} (I2I). To overcome the vagueness of text, I2I allows users to provide a reference image—such as a rough sketch, a photograph, or a 3D block-out—to guide the generation process. While I2I has different implementations (from basic diffusion over an input image to structural geometry following enabled by ControlNet) they all allow for significantly better compositional control, as the model respects the structural geometry, color palette, and lighting of the source image while applying a new style or higher level of detail.

\textit{Text-to-Video} (T2V). Video generation models attempt to generate temporal sequences directly from text. While impressive for creating short b-roll clips, generating a coherent narrative arc or complex physical interactions via text remains computationally expensive and difficult to control, since the text prompt alone is responsible for encoding color, structure, character, setting and movement instructions leaving this input modality with even stronger limits then T2I.

\textit{Image-to-Video} (I2V). This modality is increasingly preferred in professional pipelines because it separates content control from animation control. By using a high-quality, pre-approved image (often created via I2I or T2I) as the first frame, filmmakers can ensure the characters and environment are exactly as intended. The video model is then tasked only with animating that specific content. This is not only more computationally efficient as the user is not wasting time in re-rolling the video generation but also ensures that the visual identity of the asset remains stable throughout the motion sequence. Some models allow an upgraded I2V input, where both first frame and last frame control is considered, empowering the user with an even stronger content control and easier movement/animation prompting.

While the use of these models seems to challenge traditional asset production technologies, they come with many technological limits, as we mentioned before. For example, because most image and video generation systems are trained on 8-bit data—the most prevalent data type on the web (\citealp{Dornauer_2023})—these models typically produce 8-bit content. This results in a level of image quality that is notably insufficient when compared to professional Filmmaking standards. This simple fact vastly reduces the practical scope of use of asset generation models to web content and the implementation of these systems in real production workflows remains challenging.

\subsubsection{Sociomaterial aspects}\label{Sec_4.3.2}

\textit{Ad hoc coverage}. There is one significant advantage provided by direct image and video generation. In Filmmaking, one often requires brief shots for continuity or pacing, such as cut-away shots or establishing shots.\footnote{A cut-away shot is a brief insertion between two shots used for spatial orientation or to elongate a temporal beat (e.g., a character looks up and the scene cuts to a cloud). An establishing shot typically utilizes a wide angle to introduce the environment and set the scene’s context.} These elements can now be generated on demand, potentially reducing the need for additional location shooting. Editors often find themselves lacking the necessary footage for a specific transition; however, GenAI enables the creation of brief clips without the need for complex re-shoots. Furthermore, montage sequences frequently require high-impact visuals—such as an erupting volcano, an underwater dive, or extreme weather conditions like blizzards—that would be prohibitively challenging to capture physically. Through generative systems, these specific assets can be synthesized with significantly less effort.

\textit{Pipeline reconfiguration}. Much of contemporary Filmmaking relies on the digital compositing of disparate assets, many of which traditionally require substantial temporal and technical resources to generate. The integration of GenAI could allow a fundamental reconfiguration of the Filmmaking pipeline, minimizing time spent on physical production by enabling the on-demand synthesis of secondary assets. Let us consider a common scenario: a short scene requires a character to walk through a forest. There are several ways to approach this. The most traditional solution is to shoot on location in a real forest. When this location is used for only a few seconds of screen time and does not serve other scenes, the fixed costs involved become disproportionately high. To mitigate this, the scene can instead be shot on a sound stage using a green screen, with the forest environment recreated in post-production. This approach is widely used and generally effective, but it introduces a nontrivial post-production burden. The background must be digitally reconstructed, often requiring a long and monotonous asset generation phase in which trees, bushes, and other elements are modeled and rendered individually and manually by expert VFX artists. The introduction of generative systems appears to challenge this paradigm in two distinct ways, though only one proves to be genuinely cost effective. One option is to rely entirely on video generation systems to synthesize the full shot. While this can work for simple, stock-footage-like material, it becomes highly problematic when specific characters, aesthetics, or precise camera movements are required. Enforcing strict constraints on generative models typically demands specialized tooling and, even then, results are often obtained through repeated trial and error, a process commonly referred to as “re-rolling.” This significantly increases both time and cost. Moreover, even a successful synthetic shot is likely to be delivered at low resolution and 8-bit color depth, making it unsuitable for standard post-production workflows such as color grading. For this approach to be viable, all visual characteristics would need to be effectively “baked in” at generation time, which is unrealistic given that many creative decisions, most notably final color grading, are made at the very end of the pipeline. An alternative, and more practical, approach lies between these extremes. Generative systems can be used selectively for asset creation, while traditional green-screen compositing is retained for the primary subject and camera work. In this hybrid configuration, control over performance, camera motion, and scene composition remains precise, while generative tools are applied only to background elements such as trees and foliage. These assets do not require perfect fidelity, either visually or in terms of underlying data, and are often partially obscured by additional effects such as depth of field, fog, or motion blur. This midpoint solution achieves both efficiency and control, without compromising the quality or flexibility of the final output.

To conclude, asset generation techniques are apparently the most disruptive in that they promise to substitute the entire production process. Yet, given all the limits to their uses already discussed, one should rather focus on those typical problems of film production they could solve easily.

Table \ref{Tab_3} provides a synthesis of technical descriptions, reconfigurations of practices, and potential aesthetic effects for asset generation techniques.

\section{Conclusion}\label{Sec_conclusion}

This paper  has demonstrated that the integration of GenAI into Filmmaking may represent neither the replacement of human authorship nor the emergence of an autonomous creative entity. Instead, it acts as a mediator of sociomaterial reconfiguration. Through the analysis of production pipelines, it becomes evident that AI does not “create” in a vacuum; rather, it reorganizes production by transforming the physical set into a site for capturing primary subjects while delegating the generation of complex secondary assets, such as backgrounds and foliage, to algorithmic systems. This shift is not neutral: it renegotiates timeframes, costs, and professional responsibilities within the production network. On a theoretical level, our findings suggest that the co-creativity paradigm is insufficient, as it tends to decontextualize the creative act by reducing it to a dyadic, cognitive exchange between human and machine. In doing so, it overlooks the distributed agency of the wider production network and the technical constraints of professional cinematic workflows. We propose instead the framework of distributed creativity as a more fruitful interpretive lens. This perspective allows us to view agency not as a property of an individual software or person, but as a phenomenon emerging from the interactions between professionals, GenAI techniques, and pre-existing practices. AI is not a “creative partner” in the human sense, but a mediator that both enables and constrains new aesthetic practices. Finally, through the proposed taxonomy, we have shown that GenAI techniques could redefine media asset production and, consequently, the spatio-temporal dynamics of Filmmaking. Our analysis shows that while AI introduces new affordances, it follows a historical trajectory where technical constraints have consistently influenced aesthetic and professional shifts. Future research should prioritize empirical studies of these evolving workflows to further map how the category of “creative work” is being redefined. By investigating creativity in the making, we can better understand the political and aesthetic stakes of AI integration in the media industries.

\section*{Acknowledgements}

The authors consulted Gemini (Google) to assist with structural organization and editing. The authors maintain full responsibility for the final content and the original arguments presented in the paper.

\bibliographystyle{abbrvnat}
\bibliography{mybib.bib}

\newpage
\appendix
\section*{Appendix}

\begin{center}

\begin{table}[h!]
\begin{tabular}[]{ |m{2cm}|m{3cm}|m{5cm}|m{5cm}|}
     \hline
     \multicolumn{4}{|c|}{\textbf{Asset enhancement}} \\
     \hline
     
     \textbf{Technique} & \textbf{Description} & \small \textbf{Reconfiguration of practices} & \textbf{Aesthetic effects}\\
     \hline
     
     \scriptsize
     \begin{tabular}{c}
          Upscaling
     \end{tabular}
      & \scriptsize Increase of video footage resolution & \scriptsize
       \vspace{0,4cm}
       \begin{itemize}
           \item Less need to pay attention to composition in framing while shooting on set, saving time
           \item Decoupling camera resolution from final output via post-production enhancement
           \item Less need to change lenses on set to get a closer framing of the subjects: one could obtain a close-up from a long or medium shot
       \end{itemize} & \scriptsize
       \begin{itemize}
           \item New styles of découpage
           \item Different video quality standards
       \end{itemize}\\
       \hline

       \scriptsize
     Interpolation & \scriptsize Generation of new frames in-between frames with movement continuity & \scriptsize
       \begin{itemize}
           \item No need to plan slow motion in pre-production; one can do that during the editing
           \item When shooting on set for a slow motion effect one can do that at a lower frame rate with 2 benefits: less light needed on set; smaller size of the generated files (if shooting digital) or less film needed (when shooting film)
           \item Work load reduction in frame based animation
       \end{itemize} & \scriptsize
       \begin{itemize}
           \item Different quality of movements on screen
           \item Possible overabundance of slow-mo effects
       \end{itemize}\\
       \hline

       \scriptsize
     Audio denoising & \scriptsize Manipulation of audio tracks with suppression of undesired background noises & \scriptsize
       \begin{itemize}
           \item Shooting in noisy environments
           \item Usage of smaller and cheaper microphones
           \item Less need to make another take of a shot in case undesired noises were recorded but the take was overall good
       \end{itemize} & \scriptsize
       \begin{itemize}
           \item New nuances in sound design
           \item New possibilities for cinéma-vérité shootings and for Guerrilla Filmmaking
       \end{itemize}\\
       \hline

       \scriptsize
     Audio separation & \scriptsize Separation of different sounds from a single mono audio track & \scriptsize
       \begin{itemize}
           \item The roles of editor and director can merge
           \item Old audio tracks can be used for new works
       \end{itemize} & \scriptsize
       \begin{itemize}
           \item New nuances in sound design
           \item New possibilities for dubbing
       \end{itemize}\\
       \hline

       \scriptsize
     Normal/depth maps & \scriptsize Recovery of 3D from 2D video for relighting, better VFX and better color grading & \scriptsize
       \begin{itemize}
           \item New techniques for relighting and color grading
       \end{itemize} & \scriptsize
       \begin{itemize}
           \item More refined backgrounds
           \item New possibilities in the use of the depth of field
       \end{itemize}\\
       \hline
\end{tabular}

    \caption{Synthetic summary of asset enhancement techniques.}
    \label{Tab_1}
\end{table}

\end{center}

\begin{center}

\begin{table}[h!]
\begin{tabular}[]{ |m{2cm}|m{3cm}|m{5cm}|m{5cm}|}
     \hline
     \multicolumn{4}{|c|}{\textbf{Asset editing}} \\
     \hline
     
     \textbf{Technique} & \textbf{Description} & \small \textbf{Reconfiguration of practices} & \textbf{Aesthetic effects}\\
     \hline
     
     \scriptsize
     In-painting & \scriptsize Object removal and background editing & \scriptsize
       \vspace{0,4cm}
       \begin{itemize}
           \item Eliminates undesired background objects like signs or buildings when shooting outdoors. This reduces the need for location permits and enables small productions to shoot in complex environments without owner consent
           \item Removal of objects to resolve continuity errors during post-production
       \end{itemize} & \scriptsize
       \begin{itemize}
           \item New découpage choices
           \item Different video quality standards
       \end{itemize}\\
       \hline

       \scriptsize
     Out-painting & \scriptsize Reframing of static shots by enlarging the view & \scriptsize
       \begin{itemize}
           \item Expanding limited studio sets to create long shots, lowering the budget needed for building and designing physical sets
       \end{itemize} & \scriptsize
       \begin{itemize}
           \item New découpage choices
       \end{itemize}\\
       \hline

       \scriptsize
     Face transfer & \scriptsize Digital face replacement & \scriptsize
       \begin{itemize}
           \item Generating digital doubles for stunt sequences and synthetic pick-ups. It reduces physical risk for performers and eliminates the need for physical re-shoots
       \end{itemize} & \scriptsize
       \begin{itemize}
           \item New découpage choices when using stunt doubles
       \end{itemize}\\
       \hline
\end{tabular}

    \caption{Synthetic summary of asset editing techniques.}
    \label{Tab_2}
\end{table}

\end{center}

\begin{center}

\begin{table}[h!]
\begin{tabular}[]{ |m{2cm}|m{3cm}|m{5cm}|m{5cm}|}
     \hline
     \multicolumn{4}{|c|}{\textbf{Asset generation}} \\
     \hline
     
     \textbf{Technique} & \textbf{Description} & \small \textbf{Reconfiguration of practices} & \textbf{Aesthetic effects}\\
     \hline
     
     \scriptsize
     Image generation & \scriptsize Synthetic production of assets (images) & \scriptsize
       \vspace{0,4cm}
       \begin{itemize}
           \item Use of ad hoc elements for composition and frame manipulation
       \end{itemize} & \scriptsize
       \begin{itemize}
           \item Unnatural elements within the frame
       \end{itemize}\\
       \hline

       \scriptsize
     Video generation & \scriptsize Synthetic production of assets (videos) & \scriptsize
       \begin{itemize}
           \item Resolving on-set omissions via post-production synthesis of cut-away shots to avoid reshoots
           \item Synthesizing high-budget establishing shots and specialized environments (e.g., underwater sequences) without physical production costs
       \end{itemize} & \scriptsize
       \begin{itemize}
           \item Unnatural images, movements and effects
           \item Unreal transitions
           \item Non-physical photographic perspectives
       \end{itemize}\\
       \hline
\end{tabular}

    \caption{Synthetic summary of asset generation techniques.}
    \label{Tab_3}
\end{table}

\end{center}

\end{document}